\documentclass[journal]{IEEEtran}

\usepackage{cite}

\usepackage[pdftex]{graphicx}
\usepackage{amsmath}
\usepackage{todonotes}
\usepackage[hidelinks]{hyperref}
\usepackage{url}
\usepackage{amsfonts} 
\usepackage{siunitx}
\usepackage{array,multirow}
\DeclareUnicodeCharacter{2212}{\textendash}
\usepackage{newtxmath}       

\def\ii{{\hat{\imath}}}												
\def\ij{{\hat{\jmath}}}												
\def\ik{{\hat{\kappa}}}												
\begin{document}

\title{Learning Speech Emotion Representations\\ in the Quaternion Domain}

\author{Eric~Guizzo, 
Tillman~Weyde, 
Simone~Scardapane, 
and~Danilo~Comminiello~\IEEEmembership{Senior~Member,~IEEE}

\thanks{E. Guizzo and T. Weyde are with the Department of Computer Science, City, University of London, Northampton Square, London EC1V 0HB, United Kingdom. S. Scardapane and D. Comminiello are with the Department of Information Engineering, Electronics and Telecommunications (DIET), Sapienza University of Rome, Via Eudossiana 18, 00184 Rome, Italy. Corresponding author's email: 
ericguizzo@city.ac.uk.}
\thanks{This work has been performed while the first author was a PhD Visiting Student at Sapienza University of Rome, Italy.}
\thanks{This work has been partly supported by ``Progetti di Ricerca Grandi'' of Sapienza University of Rome under grant number RG11916B88E1942F.}
}


\maketitle
 
\begin{abstract}

The modeling of human emotion expression in speech signals is an important, yet challenging task.
The high resource demand of speech emotion recognition models, combined with the general scarcity of emotion-labelled data are obstacles to the development and application of effective solutions in this field.
In this paper, we present an approach to jointly circumvent these difficulties.
Our method, named RH-emo, is a novel semi-supervised architecture aimed at extracting quaternion embeddings from real-valued monoaural spectrograms, enabling the use of quaternion-valued networks for speech emotion recognition tasks.
RH-emo is a hybrid real/quaternion autoencoder network that consists of a real-valued encoder in parallel to a real-valued emotion classifier and a quaternion-valued decoder.
On the one hand, the classifier permits to optimization of each latent axis of the embeddings for the classification of a specific emotion-related characteristic: valence, arousal, dominance, and overall emotion.
On the other hand, quaternion reconstruction enables the latent dimension to develop intra-channel correlations that are required for an effective representation as a quaternion entity.
We test our approach on speech emotion recognition tasks using four popular datasets: IEMOCAP, RAVDESS, EmoDB, and TESS, comparing the performance of three well-established real-valued CNN architectures (AlexNet, ResNet-50, VGG) and their quaternion-valued equivalent fed with the embeddings created with RH-emo.
We obtain a consistent improvement in the test accuracy for all datasets, while drastically reducing the resources' demand of models.
Moreover, we performed additional experiments and ablation studies that confirm the effectiveness 
of our approach. The RH-emo repository is available at: \url{https://github.com/ispamm/rhemo}.

\end{abstract}

\begin{IEEEkeywords}
Speech Emotion Recognition, Quaternion Neural Networks, Quaternion Algebra, Transferable Embeddings
\end{IEEEkeywords}
\IEEEpeerreviewmaketitle

\section{Introduction}
\label{sec:intro}

\IEEEPARstart{H}{uman-machine} interaction is becoming increasingly important in our everyday life. 
Research on speech recognition reached near-human performance in recent years. 
Nevertheless, besides the mere sequence of words, there is  additional information that the speech can carry, in particular about emotion.
Speech Emotion Recognition (SER) is therefore acquiring a 
growing role in research on human-machine interaction, since it helps provide a more complete account of the information conveyed by speech signals. 
Despite the impressive success that neural networks have achieved in this task, SER is still challenging due to the variability of emotional expression, especially in real-world scenarios where generalization to unseen speakers and contexts is required \cite{rybka2013comparison, hozjan2003context}.
The difficulty of this task is partly related to the general scarcity of emotionally-labelled audio data, which is due to the high cost of recording and labelling such data.
Another well-known difficulty of SER is that emotional information in speech involves long-term temporal dependencies that are in the order of seconds \cite{rigoulot2013feeling, khorram2017capturing, lian2019unsupervised}. 
This forces models to analyze large temporal windows and, consequently, to use a large number of resources. 

This study proposes a joint solution for two main issues in SER research.
Broadly speaking, we propose to map speech signals into a compact multi-channel latent representation that permits having different ``emotional viewpoints'' of the signal, which are signal representations individually related to different components of human emotion, namely: valence arousal and dominance.
To this end, we make use of quaternion information processing, which is a well-established strategy to minimize models' resource demand without reducing their performance, as we discuss in detail in Sections \ref{sec:related} and \ref{sec:quaternions}.
The resulting proposed model, named Real to Emotional H-Space (RH-emo), is a semi-supervised autoencoder architecture that maps input speech signals to an embedded emotional-quaternion space. 
The axes of the embedded space are individually related to different emotion characteristics, i.e., valence, arousal, and dominance, which are represented as quaternion components.
As we will be further explored from Section \ref{sec:evaluation} onward, when used as a feature extractor that feeds into quaternion neural networks (QNNs), RH-emo improves the performance in SER tasks while considerably reducing the number of trainable parameters and computing resources, compared to equivalent real-valued models processing plain spectrograms. 
This behavior is also consistent in situations where data is very scarce.

The specific contributions of our work are the following:
%
\begin{itemize}
    \item We define a novel method, RH-emo, that draws quaternion-valued embeddings from speech signals, where each quaternion component is tailored to a specific emotional characteristic.
    \item We leverage the capabilities of quaternion emotion embeddings and the effectiveness of quaternion convolutional neural networks (QCNNs) to jointly solve two of the most significant issues related to speech emotion recognition: data scarcity and high resource demand.
    \item We extensively evaluate our approach using 4 popular SER datasets and 3 widely-used CNN-based architectures. 
    \item We provide open-source code\footnote{\url{https://github.com/ispamm/rhemo}.} and pretrained models\footnote{\href{https://drive.google.com/drive/folders/1BWvbxqnsHK7FyXB1_L_DlO6UFECkNRvz?usp=sharing}{Pretrained models: rhemo/weights}.} that can be exploited to improve the performance and efficiency of existing SER models.
\end{itemize}

The remainder of the paper is organized as follows: Section \ref{sec:related} reviews the relevant literature, Section \ref{sec:quaternions} is a brief overview of quaternion neural networks, Section \ref{sec:method} describes our proposed method in detail, Section \ref{sec:evaluation} presents our experimental setup and results, Section \ref{sec:ablation} presents the ablation studies we conduct, Section \ref{sec:discussion} discusses our outcomes and the properties of our approach and Section \ref{sec:conclusions} draws the conclusions of this paper.

\section{Background}
\label{sec:related}

In the literature, two main approaches to labeling expressed human emotions exist.
On the one hand, discrete models provide a set of fixed 
emotion categories, such as
\textit{happy, sad, angry, fearful, surprised, disgusted, neutral}. 
On the other hand, continuous models map emotions into a multidimensional 
space. 
The most common model is a 2D \textit{valence-arousal} space, where 
valence
describes the degree of emotional pleasantness and  arousal (or activation)  of the intensity of the emotion.
Dominance can be added as a third dimension describing the amount of control of a person expressing an emotion.
This encodes a so-called valence-arousal-dominance space  \cite{verma2017affect, gaertner2021multi, buechel2022emobank}.
Discrete emotions can be mapped in this continuous space although the exact mapping is not 
standardized and different studies can use slightly different mappings.

A traditional approach to SER is based on two consecutive stages: hard-coded extraction of affect-salient features followed by a learning-based classification or regression. 
Various combinations of features and classifier types have been proposed.
The most commonly used features are: base pitch, formant features, energy/spectral features, and prosody. A wide variety of classifiers has been proposed: artificial neural networks \cite{bhatti2004neural, cowie2001emotion, nicholson2000emotion}, Bayesian networks, \cite{ververidis2008fast}, Hidden Markov Models \cite{mao2009multi, nwe2003speech}, support vector machines \cite{zhou2006speech, hu2007gmm}, and Gaussian mixture models \cite{neiberg2006emotion}.
Nevertheless, in state-of-the-art methods, there is no default choice of features and classifier type \cite{el2011survey}.
With the advent of deep learning, end-to-end learning mostly replaced hard-coded feature extraction and selection, with models automatically extracting features from low-level representations of the input data (usually Fourier-based transforms, wavelet transforms, or raw audio data). 
This enables a model to fine-tune the feature extraction for a specific task and, consequently, often obtain a higher accuracy compared to engineered feature extraction. 
A range of deep learning architectures have been adopted for SER. 
The most commonly used are convolutional neural networks \cite{badshah2017speech, sun2020end, issa2020speech}, recurrent neural networks \cite{lee2015high, chernykh2017emotion} and  combinations of the two 
\cite{trigeorgis2016adieu, meyer2021improving, qamhan2020speech}
Various studies directly compare the performance of approaches using end-to-end learning and hard-coded feature extraction, showing that the former generally provides a higher classification accuracy on the same data
\cite{kim2013deep, mao2014learning, huang2014speech, han2014speech}.
Nevertheless, as a drawback, deep learning models generally require a higher computational cost and longer training times than traditional machine learning techniques and the end-to-end 
learning usually requires a large number of labelled data \cite{rossenbach2020generating, laptev2020you}. 

A well-established solution to overcome the data scarcity in SER is transfer learning by weight initialization: network weights are initialized with 
values from a network that was pretrained with a different task, possibly on a different (usually large) dataset.
Many variants of this method have been shown to improve the performance of SER models in limited-data scenarios and even when the 
task is rather distant from speech emotion \cite{macary2021use, pepino2021emotion, guizzo2021anti}.
Also, various data augmentation strategies have been successfully adopted for the same purpose, e.g. \cite{padi2020multi, shilandari2021speech}.
On the other hand, the application of dimensionality reduction transformations to the model's input data is an established strategy for reducing resource demands while limiting the loss of useful information carried by the input data.
Among others, autoencoders, PCA-based approaches, and transformer networks have been used in the field of SER \cite{fewzee2012dimensionality,patel2021impact, pepino2021emotion}, obtaining improvement both in the model's efficiency and classification accuracy.

A recent and increasingly popular strategy to improve the efficiency and the performance of deep learning models is the use of quaternion information processing \cite{tay2019lightweight, grassucci2021quaternion, grassucci2021quaternion2, grassucci2021lightweight, greenblatt2018introducing, parcollet2018quaternion, muppidi2021speech}.
Performing operations in the quaternion domain permits bootstrap intra-channel correlations in multidimensional signals \cite{bulow2001hypercomplex, mandic2010quaternion}, i.e., among the color channels of RGB-encoded images.
Moreover, due to the fewer degrees of freedom of the Hamilton product compared to the regular dot product, quaternion networks have a significantly lower number of parameters compared to the real counterparts \cite{tay2019lightweight}.
Quaternion-valued neural networks have also been successfully adopted in the audio domain \cite{comminiello2019frequency, comminiello2019quaternion} and specifically for speech recognition \cite{parcollet2018quaternion} and speech emotion recognition \cite{muppidi2021speech}.
Nevertheless, an intrinsic limitation of quaternion information processing is that it requires three or four-dimensional data as input, where intra-channel correlations exist \cite{greenblatt2018introducing, grassucci2021quaternion, grassucci2021lightweight, grassucci2021quaternion2}.
This is necessary to enable the benefits derived from the use of the Hamilton product instead of the regular dot product, as further discussed in Section \ref{sec:quaternions}.
In the audio domain, first-order Ambisonics \cite{furness1990ambisonics} signals are naturally suited for a quaternion representation, being four-dimensional and presenting strong correlations among the spatial channels,
and the application of quaternion networks to problems related to this audio format has already been extensively investigated \cite{comminiello2019quaternion, qiu2020quaternion, brignone2022efficient, grassucci2023dual}.
Nevertheless, in the vast majority of cases, audio-related machine-learning tasks deal with monaural signals, which are usually treated as 
vectors of scalars (time-domain signals), matrices of scalars (magnitude spectrograms), or 3D tensors (complex spectrograms). 
Hence they can not be naturally represented as a quaternion entity and additional processing is required to produce a suitable quaternion representation of these signals.

A number of different approaches have been proposed to overcome the necessity of having three or four-dimensional input data with intra-channel correlations.
Among others, \cite{parcollet2018quaternion} use Mel spectrograms, cepstral coefficients, and first and second-order derivatives as the four axes of the encoded quaternion.
In contrast, \cite{muppidi2021speech} convert Mel spectrograms to color-scaled images and use the RGB channels as axes of the encoded quaternion, following a computer vision-oriented approach.
Parcollet \textit{et al.} \cite{parcollet2019tele} presented two learning-based approaches to map real-valued vectors into the quaternion domain, by producing through a network four-channel representations of the input data that present meaningful intra-channel correlations.
On the one hand, the Real to H-space encoder \cite{parcollet2019tele}, applied to speech recognition tasks, consists of a simple real-valued dense layer applied at the beginning of a quaternion classifier network, which is trained jointly with the classifier.
On the other hand, the Real to H-space Autoencoder, tested in the natural language processing field (conversation theme identification) \cite{parcollet2019tele} operates in an unsupervised way. 
Such a method contains a real-valued encoder and a quaternion-valued decoder, where the latter is expected to enable both the network's embeddings and output to present meaningful intra-channel correlations that can be exploited by a quaternion-valued classifier network.

In this paper, we introduce RH-emo, a hybrid real-quaternion autoencoder-classifier architecture that is trained in a semi-supervised fashion in order to optimize each axis of the embedding dimension to different emotional characteristics: 
the first channel is optimized for discrete emotion recognition and the 3 other channels are individually optimized for the classification of valence, arousal, and dominance (as shown in Figure \ref{fig:diagram}). 
RH-emo is intended to be used as a feature extractor that permits using QNNs for SER tasks with real-valued signals without additional 
preprocessing.
This approach has two advantages: it improves the performance of SER models even in situations where data is scarce and
it drastically reduces the number of network parameters, consequently reducing the resource demand.
We extend the approach of the quaternion autoencoder in \cite{parcollet2019tele} by specializing the learned quaternion representation for our specific task (SER), where the different axes are optimized for the detection of different emotional characteristics that are coherent with the most used criteria of emotion classification.
Moreover, we implement it with a more complex architecture (deep convolutional autoencoder) and we apply it to a different domain: emotion recognition from speech audio.

\section{Quaternion convolutional neural networks}
\label{sec:quaternions}
 Operations between quaternion numbers are defined in the quaternions algebra $\mathbb{H}$.
A quaternion Q is a four-dimensional extension of a complex number, defined as 
$\mathbf{q} = q_0 + q_1 \ii + q_2 \ij + q_3 \ik = q_0 + q$,
where, $q_0$, $q_1$, $q_2$ are real numbers, and $\ii$, $\ij$ and $\ik$ are the quaternion unit basis. 
In this representation $q_0$ is the real part and $q_1 \ii + q_2 \ij + q_3 \ik$ is the imaginary part, where $\ii^{2} = \ij^{2} = \ik^{2} = -1$ and $\ii \ij = - \ij \ii$.
From the latter assumption follows that the quaternion vector multiplication is not commutative.
A quaternion can also be represented as a matrix of real numbers:

\begin{equation}
\label{eq:qnumbmat}
    \mathbf{q} = \left[ {\begin{array}{*{20}c}
   \hfill {q_0} & \hfill { - q_1 } & \hfill { - q_2 } & \hfill { - q_3 } \\
   \hfill {q_1 } & \hfill {q_0 } & \hfill { - q_3 } & \hfill {q_2 } \\
   \hfill {q_2 } & \hfill {q_3 } & \hfill {q_0 } & \hfill { - q_1 } \\
   \hfill {q_3 } & \hfill { - q_2 } & \hfill {q_1 } & \hfill {q_0 } \\
\end{array}} \right].
\end{equation}

Analogously to real and complex numbers, a set of operations can be defined in the quaternion space: 
\begin{itemize}
    \item \textbf{Addition}: $\mathbf{q}+\mathbf{p} = (q_0 + p_0) + (q_1 + p_1)\ii + \\ (q_2 + p_2)\ij + (q_3 + p_3)\ik$
    \item \textbf{Conjugation}: $\mathbf{q}^{*} = q_0 - q_1 \ii − q_2 \ij - q_3 \ik$
    \item \textbf{Scalar multiplication}: $\lambda \mathbf{q} = \lambda q_0 + \lambda q_1 \ii+ \lambda q_2 \ij + \lambda q_3 \ik$
    \item \textbf{Element multiplication} (or \textbf{Hamilton product}):
    \begin{equation}
	\begin{split}
		\mathbf{q} \otimes \mathbf{p} &= \left(q_0 + q_1\ii + q_2\ij + q_3\ik\right)\left(p_0 + p_1\ii + p_2\ij + p_3\ik\right) \\
		&= \left(q_0 p_0 - q_1 p_1 - q_2 p_2 - q_3 p_3\right) \\
		&+ \left(q_0 p_1 + q_1 p_0 + q_2 p_3 - q_3 p_2\right)\ii \\
		&+ \left(q_0 p_2 - q_1 p_3 + q_2 p_0 + q_3 p_1\right)\ij \\
		&+ \left(q_0 p_3 + q_1 p_2 - q_2 p_1 + q_3 p_0\right)\ik.
	\end{split}
	\label{eq:hamilton}
\end{equation}
\end{itemize}


The quaternion convolutional neural network (QCNN) is an extension of the real-valued convolutional neural network to the quaternion domain. 
For each input vector of a quaternion layer, the dimensions are split into four parts to compose a quaternion representation.
In a quaternion-valued fully-connected layer
the parameters matrices are treated as a single quaternion entity with four components,
even though they are manipulated as matrices of real numbers \cite{gaudet2018deep}.
In a quaternion layer, the dot product operations used in real layers are replaced with the Hamilton product (eq.~\eqref{eq:hamilton}) between the input vector and a quaternion-represented weight matrix.
This allows the processing of all input channels together as a single entity maintaining original intra-channels dependencies because the weights submatrices are shared among the input channels.
Consequently, quaternion layers permit to spare the 75\% of free parameters compared to their real-valued equivalents because, as shown in eq.~\eqref{eq:hamilton}, the same components are re-used to build the output matrix.

In a QCNN, the convolution of a quaternion filter matrix with a quaternion vector is performed as the Hamilton product between the real-valued matrices representation of the input vector and filters. 
A quaternion convolution between a quaternion input vector 
$\mathbf{x} = x_0 + x_1 \ii + x_2 \ij + x_3 \ik$
and a quaternion filter 
$ W = W_0 + W_1 \ii + W_2 \ij + W_3 \ik$
can be defined as:

\begin{equation}
\label{eq:qprod}
    W * x = \left[ {\begin{array}{*{20}c}
   \hfill {W_0} & \hfill { - W_1 } & \hfill { - W_2 } & \hfill { - W_3 } \\
   \hfill {W_1 } & \hfill {W_0 } & \hfill { - W_3 } & \hfill {W_2 } \\
   \hfill {W_2 } & \hfill {W_3 } & \hfill {W_0 } & \hfill { - W_1 } \\
   \hfill {W_3 } & \hfill { - W_2 } & \hfill {W_1 } & \hfill {W_0 } \\
\end{array}} \right] * \left[ {\begin{array}{*{20}c}
   {x_0 } \hfill  \\
   {x_1 } \hfill  \\
   {x_2 } \hfill  \\
   {x_3 } \hfill  \\
\end{array}} \right].
\end{equation}

The optimization of quaternion-valued networks is identical to the one of a real network and can be achieved through regular backpropagation.
This is possible because of the use of split activation and loss functions, as introduced in \cite{parcollet2019tele, ujang2011quaternion}.
These functions map a quaternion-like entity back to the real domain, consequently enabling the use of standard loss functions for the network training.

\section{The Proposed RH-emo Model}
\label{sec:method}

\subsection{Approach} 
\label{subsec:approach}

The main aim of RH-emo is to map real-valued spectrograms to the quaternion domain, building compact emotion-related quaternion embeddings where each axis is optimized for a different emotional characteristic.
In the embedded dimension, the real axis of the quaternion is optimized for the discrete classification of 4 emotions: \textit{neutrality, anger, happiness, sadness} and the 3 complex axes are optimized for the prediction of emotion in a \textit{valence, arousal} and \textit{dominance} 3D space.
This representation exploits the natural predisposition of quaternion algebra to process data where a 4 or 3-channels representation is meaningful.
Nevertheless, in most machine learning applications of quaternion algebra, the input data is naturally organized with a meaningful shape, as happens for instance with RGB/RGBA images (where the color/alpha channels are treated as different quaternion axes) and first-order Ambisonics audio signals (where the 4 spatial channels are considered as the quaternion axes).
In our case, instead, such quaternion representation is created through a semi-supervised learning procedure, where the different axes are forced to contain information related to different complementary emotion characteristics. 
Therefore, in a certain sense, the axes of this embedded dimension can be thought of as different ``emotional points of view" of an audio signal.

RH-emo is intended to be used as a pretrained feature extractor to enable the use of quaternion-valued neural networks for SER tasks applied to monoaural audio signals.
On the one hand, the emotion-related disentanglement among channels helps to enhance the performance of SER models, especially under conditions of data scarcity.
Whereas, on the other hand, the reduced dimensionality together with the enabled possibility to classify the data with quaternion-valued networks permits to spare of a large number of network parameters, consequently lowering the resource demand and speeding up the training.

\subsection{RH-emo Architecture} 
\label{subsec:architect}

\begin{figure}[!tb]
\begin{minipage}[b]{1.0\linewidth}
  \centerline{\includegraphics[width=8.5cm]{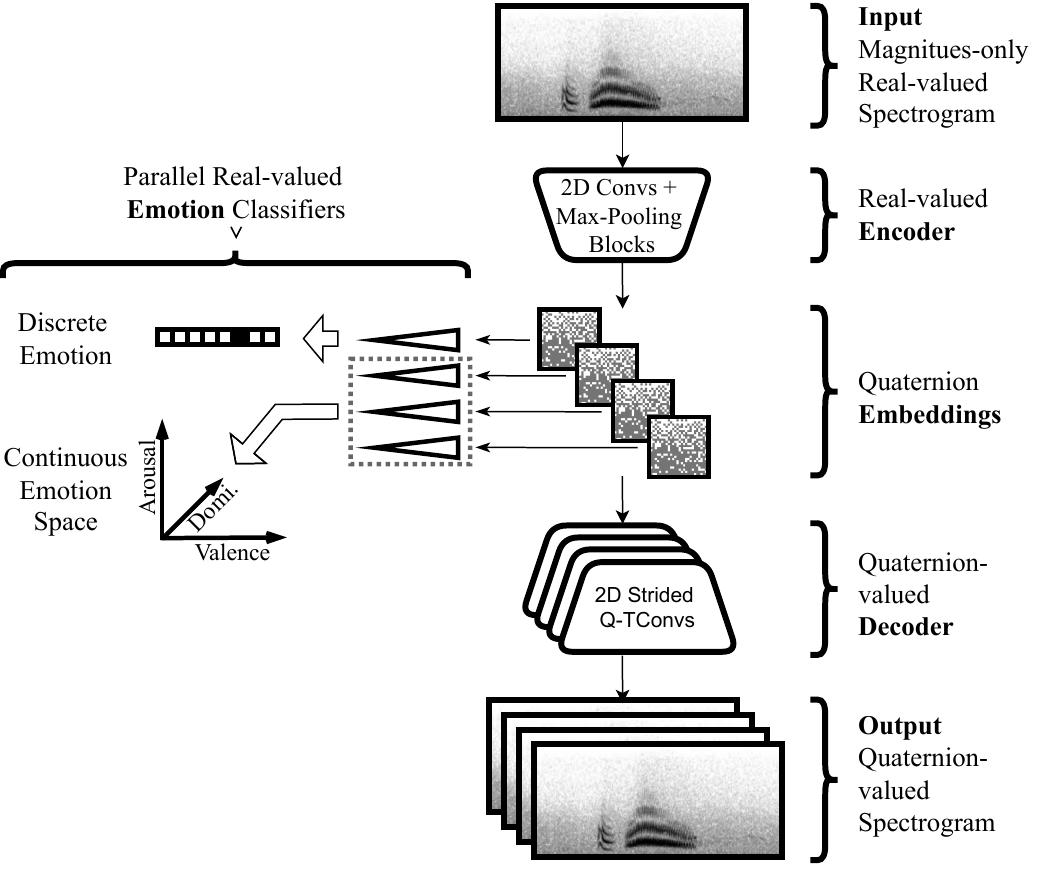}}
\end{minipage}
\caption{RH-emo Block Diagram. An input magnitudes-only spectrogram is first propagated into a real-valued convolutional encoder that generates embeddings with a [4x64x64] shape. The network is then split into two branches: a completely unsupervised quaternion-valued decoder that reconstructs the input spectrogram projecting it in a four-channel quaternion space and a set of 4 parallel real-valued supervised classifiers, each connected to one of the four channels of the embeddings and separately classifying different emotion characteristics: discrete emotion, valence, arousal, and dominance.}
\label{fig:diagram}
\end{figure}

RH-emo is a hybrid real/quaternion autoencoder network. 
Its structure is similar to R2Hae \cite{parcollet2019tele}, nevertheless, RH-emo is based on a convolutional design and it embraces multiple classification branches, as opposed to R2Hae. 
We used a public PyTorch implementation of convolution layers and operators\footnote{https://github.com/Orkis-Research/Pytorch-Quaternion-Neural-Networks}.
As Figure \ref{fig:diagram} shows, our RH-emo is composed of three components: an encoder $E(X)$ acting on the (real-valued) input spectrogram, producing an embedded vector. 
The output of the encoder is then fed separately to a (quaternion-valued) decoder $D(Z)$ to reconstruct the original spectrogram and to a classification head $C(Z)$ for performing emotion recognition. 
The classifier outputs four separate predictions $y_D$, $y_v$, $y_a$, and $y_d$ which are, respectively, a discrete and a continuous (in the valence, arousal, dominance space) categorization of the emotional content of the spectrogram. 
The specific architecture for each of these blocks, as well as the loss function we optimize and the training strategy we adopt, is described more in detail in the following paragraphs.

\subsubsection{Encoder}

The input data, a magnitudes-only real-valued spectrogram in our case, is forward propagated through a real-valued autoencoder made up of 3 convolution blocks. 
Each block contains a 2D convolution layer (ReLU activations, 3x3 kernels, single-pixel stride, increasing channels number: 1, 2, 4), followed by max-pooling layers of dimension [2x2], [2x1], [2x1]. 
Moreover, only between the first and the second block, a batch normalization layer is present.
The encoder produces an embedded vector that presents a dimensionality reduced by a factor of 0.25 compared to the input.
In our experiments, we use input spectrograms with a shape of 1x512x128 (channels, time-steps, frequency-bins) and the embedded dimension created by the encoder has a shape of 4x64x64.
The embedded vector is then  forward propagated in parallel into four distinct real-valued classifiers and also into a quaternion-valued decoder. 
It is therefore important that the embedded vector contains a number of elements that is multiple of four, in order to be properly treated as a quaternion by the decoder section of the network.

\subsubsection{Classifiers}

Each classifier consists of a sequence of 3 real-valued fully connected layers, where the first 2 contain 4096 neurons and are followed by a dropout layer.
In the first classifier, the output layer contains 4 output neurons (the number of emotional classes to be classified) and softmax activation.
Instead, the other 3 classifiers are identical and have one single output neuron with sigmoid activation, as they are individually aimed at a binary classification task: the prediction of ``high'' or ``low'' valence, arousal, and dominance, respectively.

\subsubsection{Decoder}

The decoder mirrors the encoder's structure but uses quaternion-valued 2D transposed convolutions with a stride that mirrors the pooling dimensions of the encoder, instead of the sequence of 2D real-valued convolutions and 2x2 max-pooling and a quaternion-valued batch normalization layer instead of its real-valued counterpart.
The output of the decoder is therefore a matrix with the same dimensions as the input, but with 4 channels instead of a single one.

\subsection{Loss Function}
\label{subs:loss}

The loss function we minimize during the training of RH-emo is a weighted sum of the binary crossentropy reconstruction loss between the input spectrogram and the decoder's output, the categorical crossentropy classification loss of the emotion labels predicted by the supervised classifier in the middle of the network (discrete, valence and arousal).

The objective function we minimize is, therefore:
\begin{equation}
\begin{split}
    {\cal{L}} &= \text{BCE}_(X, Y_r) +  \beta \cdot \{\text{CE}_(p,t) \\
    &+ \alpha \cdot [\text{BCE}_(v_p, v_t) + \text{BCE}_(a_p, a_t) + \text{BCE}_(d_p, d_t)]\}
\end{split}
\label{eq:loss}
\end{equation}
where $BCE$ is the binary crossentropy loss, $CE$ is the categorical crossentropy loss, $\beta$ and $\alpha$ are scalar weight factors, $X$ is the input spectrogram, $Y_r$ is the decoder's output re-mapped to the real domain through the split activation function (as discussed below), $p$ and $t$ are respectively the discrete emotion prediction and truth label, $v_p$/$v_t$, $a_p$/$a_t$ and $d_p$/$d_t$ are respectively the valence, arousal and dominance prediction, and truth labels.

For the reconstruction loss computation, it is necessary to map the quaternion-valued decoder output back to the real domain, in order to have the same shape as the input vector.
For this purpose we use a stratagem similar to the ``split activation'' described in \cite{parcollet2019tele, ujang2011quaternion}: we perform an element-wise mean across the channel dimension of the quaternion output, bringing back the 4-channels vector to a single-channel shape.
During the training, this forces the model to not weigh the intra-channel correlations among the quaternion axes in the reconstruction term of the loss (the leftmost term of eq.~\eqref{eq:loss}).
Our expectation is that this leaves room for the emotion recognition term of the loss (the rightmost term of eq.~\eqref{eq:loss}) for tuning these correlations, making them related to the emotional information.

\subsection{Training strategy}
\label{subsec:training}

For the RH-emo training, we use the Interactive Emotional Dyadic Motion Capture Database (IEMOCAP) dataset \cite{busso2008iemocap}, which includes: 5 speakers, 7529 utterances, 9:32 hours of audio, 10 emotion labels and it is in the English language. 
We selected this specific dataset for the following reasons: it is one of the most popular SER datasets, it contains a large number of datapoints, it is not limited to a restricted set of sentences, emotions are expressed by actors with a natural feeling rather than being over-emphasized \cite{busso2008iemocap}
and it is labelled both in the discrete and continuous (valence, arousal, dominance) emotional domains.

We apply 4 preprocessing stages to the raw data: we first extract 4-second non-overlapped fragments (or zero-pad if a datapoint is shorter that this duration). Then, we compute the short-time Fourier transform (STFT) using 16 ms sliding windows with 50\% overlap, applying a Hamming window and discarding the phase information. 
After this point, we normalize the whole dataset between 0 and 1 and, in the end, we zero-pad the spectrograms to match a shape of 512 (time-steps) x 128 (frequency-bins).

To permit proper convergence, we perform the training in 2 consecutive stages: we first train the network until convergence with the $\beta$ weight set to 0.
This removes the rightmost term from eq.~\eqref{eq:loss}, consequently eliminating the emotion classification part of the loss.
Doing so, we train the network in a completely unsupervised way only to perform a quaternion projection of the real input spectrogram, without taking into account any emotion-related information.
After this stage, we re-train the network adding also the classification term in the loss in order to specialize the learned representations to the emotion recognition task, but also maintaining the embedded vector in a quaternion-compatible shape that is meaningful for the decoder part of the network.
For this stage, we performed a grid search to find the best combination of the emotion classification weights  $\beta$ and  $\alpha$ and we ended up using $\beta=0.01$ and $\alpha=100$.
This means that overall we weigh more the reconstruction error in the loss function (thanks to the low $\beta$), and we weigh more the dimensional emotion classification compared to the discrete classification (thanks to the high $\alpha$).

While for the first, completely unsupervised, training stage we use all data available with IEMOCAP, in the second supervised stage we use only a subset of the dataset, including only the datapoints related to 4 emotions (\textit{angry, happy, neutral, sad)} and we merge the classes \textit{happy} and \textit{excited} as one single emotion class \textit{happy}.
This is a standard procedure with IEMOCAP, as the other labels contained in the dataset are highly imbalanced.
For both training stages, we use subsets of approximately 70\% of the data for training, 20\% for validation, and 10\% for the test set. 
We use a learning rate of 0.001 in the first stage and of 0.000001 in the second one, a batch size of 20 and the Adam optimizer \cite{kingma2014adam}. 
We use dropout at 50\% in the classification branches for the second training stage.
We apply early stopping by testing at the validation loss improvement with patience of 100 epochs in the first stage and 30 epochs for the second one.

After these 2 training stages, we obtain a test reconstruction loss (the isolated leftmost term of eq.~\eqref{eq:loss}) of 0.00413 and competitive test classification accuracy: 60.7\% for the discrete classification and respectively 65.4\%, 75.3\% and 70.2\% for the valence, arousal, and dominance dimensions.

\section{Evaluation}
\label{sec:evaluation}

In order to test the capabilities and properties of RH-emo, we compare the classification accuracy for SER tasks obtained with real-valued CNN networks and equivalent quaternion-valued versions of them (QCNNs).
For the quaternion versions we keep the same architecture of the real CNNs, but we use quaternion-valued convolution and quaternion-valued fully connected layers instead of the canonical real-valued ones, with the exception of the final layer of the networks, which are real-valued also in the QCNNs.
For the real networks, we use the magnitudes-only spectra as input, while for the quaternion networks we use the embeddings generated with RH-emo pretrained on IEMOCAP.
Moreover, we compare and combine our approach with a standard transfer learning method performed on the same dataset (IEMOCAP): pretraining with weight initialization.
Therefore we have two distinct types of pretraining: the pretraining of the RH-emo network, which we use to compute the emotional embeddings, and the pretraining of the CNNs that we use to perform the actual SER task. 
Both pretrainings are performed on the IEMOCAP dataset.
To avoid confusion, from here on we will refer to the first as RH-emo pretraining and to the latter as CNN's pretraining.

\begin{figure}[!tb]
    \centerline{\includegraphics[width=0.46\textwidth]{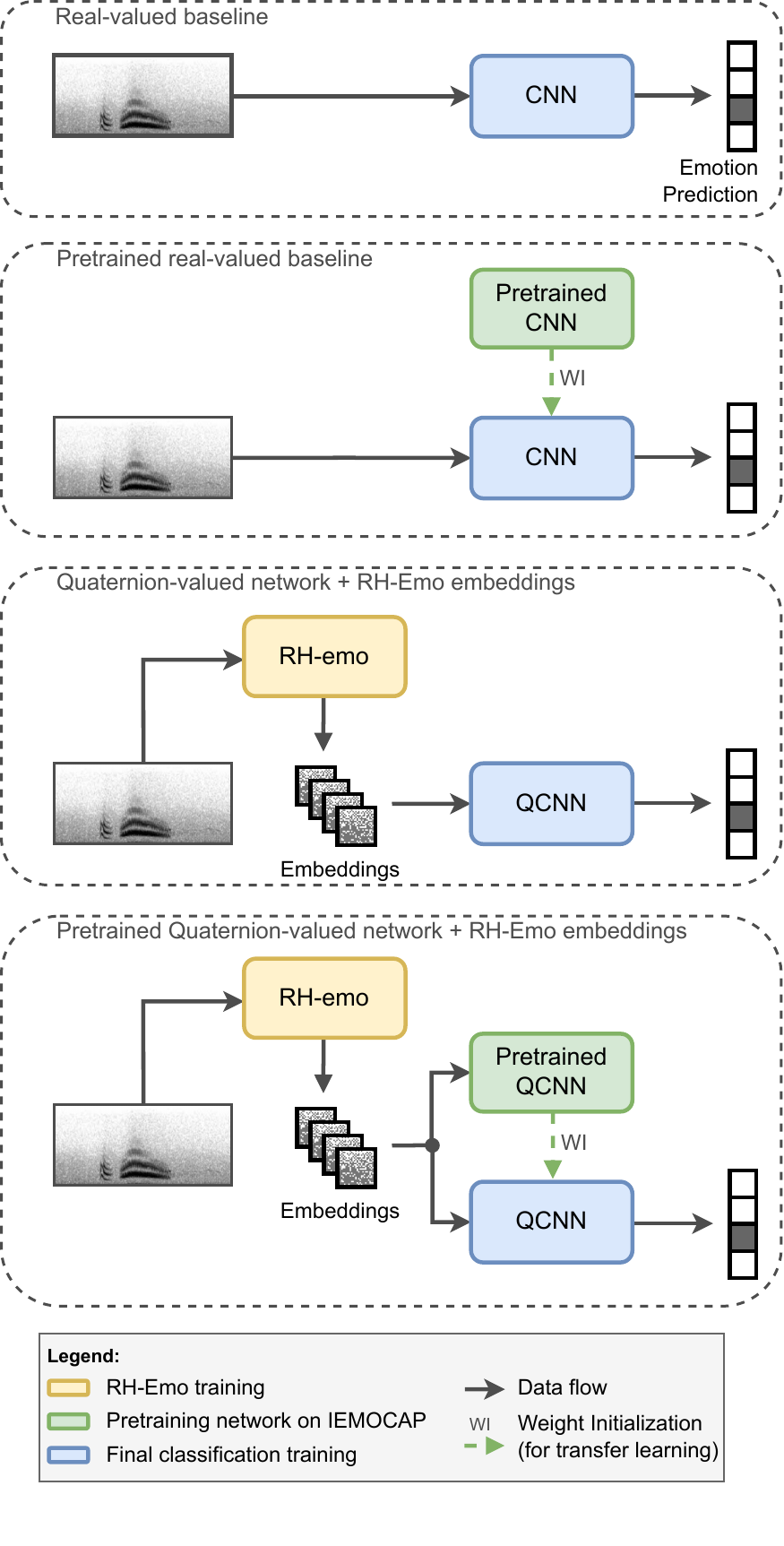}}
\caption{Block diagram of our experimental setup. The yellow-to-blue color coding reflects 3 consecutive training stages. There are 2 separate pretraining stages: RH-emo pretraining (yellow) and CNN pretraining (green). 
The straight arrows indicate the data flow, while the dotted arrows, accompanied by the word WI, show where the weights of a pretrained network are used to initialize the initial weights of an identical network (transfer learning).
The real-valued baseline is a regular CNN with random weight initialization, upper row. The pretrained real-valued baseline is the same network, but its weights are initialized with the ones of an identical network pretrained on IEMOCAP (the same dataset used to train RH-emo), second row. The quaternion-valued network is a quaternion-valued version of the real-valued baselines, in which (4 channel) input is generated by forward propagating the input spectrogram in RH-emo's encoder, third row. 
The pretrained quaternion-valued network is identical to the latter, but the weights of the CNN are initialized with the ones of an identical network pretrained on IEMOCAP, last row.
}

\label{fig:experimental_setup}
\end{figure}

Figure \ref{fig:experimental_setup} depicts all cases we include in our experimental setup.
The color coding of Figure \ref{fig:experimental_setup} shows the 3 consecutive stages of our experiments: first, we pretrain RH-emo (yellow), then we pretrain the CNNs (orange) on IEMOCAP and finally we train or retrain the CNNs on other datasets.
We have two types of baseline: the first one, shown in the upper row of Figure \ref{fig:experimental_setup}, is a standard real-valued CNN with randomly-initialized weights.
As a further baseline, as depicted in the second row of Figure \ref{fig:experimental_setup}, we test a standard transfer learning approach applied to the real-valued CNNs: we pretrain on IEMOCAP (the same dataset used to train RH-emo) and we then initialize all weights of the SER CNNs but the ones of the final classification layer.
The last two rows of Figure \ref{fig:experimental_setup}, instead, show our approach, where we use RH-emo as a feature extractor to feed quaternion-valued CNNs.
In the third row, only RH-emo pretraining happens, while in the last row both RH-emo and CNNs pretraining are performed.
In the latter case, we first pretrain RH-emo, then we pretrain the CNN on IEMOCAP, and finally, we re-train the same CNN on different datasets.

\subsection{Experimental Setup}
\label{subsec:experimentalsetup}

We evaluate RH-emo with 3 benchmark SER datasets:
\begin{enumerate}
  \item RAVDESS, the Ryerson Audio Visual Database of Emotional Speech and Song \cite{livingstone2018ryerson}. 24 speakers, English language, 2542 utterances, 2:47 hours of audio, 8 emotion labels. 
  \item EmoDB, a Database of German Emotional Speech  \cite{burkhardt2005database}. 10 speakers, German language, 535 utterances, 25 min of audio, 7 emotion labels.
  \item TESS, the Toronto Emotional Speech Set \cite{dupuis2011recognition}. 2 speakers, English language, 2800 utterances, 1:36 hours of audio, 7 emotion labels.
\end{enumerate}

The preprocessing pipeline for these datasets is identical to the one we applied to IEMOCAP, as described in Section \ref{sec:method}, except for the final normalization step. 
For the quaternion-valued networks we normalize data between 0 and 1 (as required by RH-emo), and for the real-valued networks we normalize to 0 mean and unity standard deviation to permit proper convergence. 

We apply this approach to 3 popular CNN architectures with increasing capacity: VGG16 \cite{simonyan2014very}, AlexNet \cite{krizhevsky2012imagenet} and ResNet-50 \cite{he2016deep}, based on the Torchvision implementations\footnote{\url{https://pytorch.org/vision/stable/_modules/torchvision.html}}.
These implementations present an adaptive average pooling layer between the convolution-based feature extractor and the fully-connected classifier.
This permits to obtain an identical output shape from the feature extractor for any input dimension.
We removed this layer from only VGG16, in order to test the behavior of our approach also in this situation.
Doing this, in fact, the feature extractor presents a reduced output dimensionality when the networks are fed with the quaternion embeddings (75\% smaller than using the real spectrograms), enabling to spare of a major number of network parameters.

\begin{table}[!t]
\caption{Pretraining results on IEMOCAP}
\label{table:iemocap}
\centering
\begin{tabular}{c|c|c|c|c}
\hline
\hline
\textbf{Arch.} & \textbf{Method} & \textbf{Params} & \textbf{Train acc.} & \textbf{Test acc.}\\

\hline
{RH-emo}  & / & \num{1.3e8} & 80.34 & 60.7 \\
\hline
\multirow{2}{*}{VGG16}  & Real & \num{1.6e8} & 74.88 & 62.87 \\
 & \textbf{RH-emo+Quat} & \num{1e7} & 72.25 & 71.10 \\

\hline
\multirow{2}{*}{AlexNet}  & Real & \num{5.7e7} & 71.02  & 63.33 \\
  &  \textbf{RH-emo+Quat} & \num{1e7} & 71.81 & 70.31 \\

\hline
\multirow{2}{*}{ResNet}  & Real & \num{2.3e7} & 61.05 & 57.20 \\
 &  \textbf{RH-emo+Quat} & \num{4.9e6} & 73.03 & \textbf{71.20} \\

\hline
\hline
\end{tabular}
\end{table}
\setlength{\tabcolsep}{1.5pt}
\begin{table}[!t]
\caption{Results for RAVDESS}
\label{table:ravdess}
\centering
\begin{tabular}{c|c|c|c|c|c}
\hline
\hline
\textbf{Arch.} & \textbf{Method} & \textbf{Params} & \textbf{Train acc.} & \textbf{Test acc.} & \textbf{Test UAR}\\

\hline
\multirow{4}{*}{VGG16}  & Real & \num{1.6e8} & 47.10 & 41.06 & 40.07\\
 & \textbf{RH-emo+Quat} & \num{1e7} & 55.50 & 49.85 & \textbf{48.28}\\
 & Real-Pre & \num{1.6e8} & 67.86 & 45.30 & 46.33\\
 & \textbf{RH-emo+Quat-Pre} & \num{1e7} & 67.08 & 53.79 & 46.88\\

\hline
\multirow{4}{*}{AlexNet}  & Real & \num{5.7e7} & 54.55  & 46.36 & 36.36\\
  &  \textbf{RH-emo+Quat} & \num{1e7} & 50.62 & 43.94 & 38.21\\
  & Real-Pre & \num{5.7e7} & 83.54 & 51.06 & 45.71 \\
   & \textbf{RH-emo+Quat-Pre} & \num{1.4e7} & 63.16 & 47.58 & 41.29\\
  
\hline
\multirow{4}{*}{ResNet}  & Real & \num{2.3e7} & 72.84 & 43.48 & 33.16\\
 &  \textbf{RH-emo+Quat} & \num{4.9e6} & 91.29 & \textbf{55.15} & 46.51\\
 & Real-Pre & \num{2.3e7} & 22.16 & 18.79 & 13.33\\
  & \textbf{RH-emo+Quat-Pre} & \num{4.9e6} & 89.54 & 52.42 & 44.27\\

\hline
\hline
\end{tabular}
\end{table}

\begin{table}[!tb]
\caption{Results for EmoDB}
\label{table:EmoDB}
\centering
\begin{tabular}{c|c|c|c|c|c}
\hline
\hline
\textbf{Arch.} & \textbf{Method} & \textbf{Params} & \textbf{Train acc.} & \textbf{Test acc.} & \textbf{Test UAR}\\

\hline
\multirow{4}{*}{VGG16}  & Real & \num{1.6e8} & 72.74 & 70.00 & 58.86\\
 & \textbf{RH-emo+Quat} & \num{1e7} & 79.54 & 50.00 & 41.73\\
 & Real-Pre & \num{1.6e8} & 78.16 & 52.00 & 46.95\\
 & \textbf{RH-emo+Quat-Pre} & \num{1e7} & 75.00 & 47.00 & 40.11\\

\hline
\multirow{4}{*}{AlexNet}  & Real & \num{5.7e7} & 63.1  & 47.00 & 40.77\\
  &  \textbf{RH-emo+Quat} & \num{1e7} & 82.3 & 49.00 & 41.99\\
  & Real-Pre & \num{5.7e7} & 71.45 & 67.00 & 59.93\\
   & \textbf{RH-emo+Quat-Pre} & \num{1.4e7} & 77.63 & 71.00 & 63.89\\
  
\hline

\multirow{4}{*}{ResNet}  & Real & \num{2.3e7} & 99.47 & 48.00 & 42.76\\
 &  \textbf{RH-emo+Quat} & \num{4.9e6} & 99.73 & \textbf{73.00} & \textbf{65.64}\\
 & Real-Pre & \num{2.3e7} & 100.00 & 72.00 & 64.04 \\
  & \textbf{RH-emo+Quat-Pre} & \num{4.9e6} & 99.73 & 46.00 & 38.34\\

\hline
\hline
\end{tabular}
\end{table}

For all experiments we used a learning rate of 0.00001, ADAM optimizer, and a batch size of 20 samples, we apply early stopping with the patience of 20 epochs on the validation loss and we split the training, validation, and test sub-sets with approximately 70\%, 20\% and 10\% of the data, respectively.

The main aim of this research is to provide a valid comparison between the proposed approach (quaternion-valued CNNs fed with RH-Emo embeddings) and standard equivalent real-valued architectures, isolating as much as possible the pure difference between them. 
We configured our experimental setup in order to show the performance difference between real and corresponding quaternion CNNs fed with the emotional quaternion embeddings.
Therefore, we paid attention to performing each experiment in as-close-as-possible conditions, rather than optimizing each architecture for each different dataset, in order to highlight the properties of our approach. 
State-of-the-art results for SER tasks usually involve more complex solutions, as, among others, data augmentation \cite{pham2021hybrid, jothimani2022mff, etienne2018speech, xu2021speech}, attention \cite{pham2021hybrid, xu2021speech, ho2020multimodal,kakouros2022speech}, adversarial attacks \cite{latif2018adversarial}, multimodal processing \cite{ho2020multimodal, bouali2022cross}, speaker-aware processing \cite{kim2021emoberta, li2020hitrans}, transformer designs \cite{ho2020multimodal, li2020hitrans}. 
Moreover, the state-of-the-art approach can be radically different for each dataset, and therefore using the best method for each dataset would not permit having the same configuration for all possible aspects in both RH-Emo experiments and the baselines. 
This would add much more complexity to the setup, consequently making it less straightforward to isolate and understand the properties of our approach.

Because of these reasons and the fact that many existing studies are based on different methods to compute the scores, different data splits and may use multiple data domains, our results can not be directly compared to the current state-of-the-art accuracy for these datasets, which, to the best of our knowledge are 75.60\% for IEMOCAP \cite{kakouros2022speech}, 87.5\% for RAVDESS \cite{bouali2022cross}, 88.47\% for EmoDb \cite{pham2021hybrid} and 99.6\% for TESS \cite{jothimani2022mff}.

\begin{table}[!tb]
\caption{Results for TESS}
\label{table:TESS}
\centering
\begin{tabular}{c|c|c|c|c|c}
\hline
\hline
\textbf{Arch.} & \textbf{Method} & \textbf{Params} & \textbf{Train acc.} & \textbf{Test acc.} & \textbf{Test UAR}\\

\hline
\multirow{4}{*}{VGG16}  & Real & \num{1.6e8} & 99.54 & 97.62 & 98.51\\
 & \textbf{RH-emo+Quat} & \num{1e7} & 98.87 & 97.62 & 96.67\\
 & Real-Pre & \num{1.6e8} & 99.95 & 99.52 & 98.95\\
 & \textbf{RH-emo+Quat-Pre} & \num{1e7} & 98.72 & 97.85 & 97.92\\

\hline
\multirow{4}{*}{AlexNet}  & Real & \num{5.7e7} & 99.18  & 98.01 & 97.03\\
  &  \textbf{RH-emo+Quat} & \num{1e7} & 99.54 & 98.56 & 97.34\\
  & Real-Pre & \num{5.7e7} & 100.00 & 98.01 & 98.95\\
   & \textbf{RH-emo+Quat-Pre} & \num{1.4e7} & 99.75 & 98.81 & 97.38\\
  
\hline
\multirow{4}{*}{ResNet}  & Real & \num{2.3e7} & 100.00 & 97.38 & 97.84\\
 &  \textbf{RH-emo+Quat} & \num{4.9e6} & 100.00 & \textbf{99.76} & \textbf{99.58}\\
 & Real-Pre & \num{2.3e7} & 59.88 & 57.53 & 56.72\\
  & \textbf{RH-emo+Quat-Pre} & \num{4.9e6} & 100.00 & 99.28 & 97.91\\

\hline
\hline
\end{tabular}
\end{table}

\setlength{\tabcolsep}{8.8pt}

\begin{table}[!t]
\caption{Test Accuracy Results}
\label{table:statistics}
\centering
\begin{tabular}{c|c|c|c|c}
\hline
\hline
\multirow{2}{*} {\textbf{Dataset}} & \multicolumn{3}{c|}{\textbf{Average improvement}} &  {\textbf{Best}}\\
\cline{2-4}
 & \textbf{No pret.} & \textbf{Pret.} & \textbf{Overall} & \textbf{improvement}\\

\hline
{IEMOCAP}  & 9.74 & / & / & 7.87 \\
\hline
RAVDESS  & 6.01 & 12.88 & 9.45  & 4.09 \\
\hline
EmoDB  & 2.34 & -9.00 & -3.34 & 1.00 \\
\hline
TESS & 0.97 & 13.63 & 7.30 & 0.24\\

\hline
\hline
\end{tabular}
\end{table}

\subsection{Experimental Results}
\label{subsec:experimentalrtesults}

Table \ref{table:iemocap} shows the pretraining results we obtained on IEMOCAP, while Tables \ref{table:ravdess}, \ref{table:EmoDB}, and \ref{table:TESS} provide the results on RAVDESS, EmoDB, and TESS, respectively. 
Table \ref{table:statistics}, shows the average and best test accuracy improvement provided by our approach, among all CNN architectures for each dataset. 
Here, average improvement refers to the difference between the average test accuracy among all real-valued and all quaternion-valued outcomes, whereas the best improvement is the difference between the best real-valued and the best quaternion-valued accuracy we obtained.
For the core results (Tables \ref{table:ravdess}, \ref{table:EmoDB}, and \ref{table:TESS}) we include also the test set results in terms of Unweighted Average Recall (UAR).
This gives further insight into the model's generalization performance with a metric that does not take into account possible imbalance of the datasets' labels.

The results clearly show that our approach enhances the model's performance while improving its efficiency.
For all datasets, the quaternion CNNs fed with RH-emo embeddings provide the best test accuracy overall, with an accuracy improvement of 6.01 percentage points (pp) for RAVDESS, 2.34 pp for EmoDB, and 0.97 for TESS in the case we do not apply CNN pretraining. 
The only case where our approach does not improve the test accuracy is with the EmoDB dataset, applying CNN pretraining, where we have a performance drop of 9 pp.
In the other cases where we applied CNN pretraining, our approach provides a strong average improvement of 12.88 and 13.63 pp, respectively for RAVDESS and TESS.
Moreover, the test set results in terms of UAR metric confirm the overall trend of the accuracy metric. 
Nevertheless, in one single case (VGG-16 network on RAVDESS) there is a narrow inconsistency between the two metrics. 
Here the pretrained QCNN shows the best test accuracy, while the best UAR score is given by the non-pretrained QCNN.

The results computed on IEMOCAP (Table \ref{table:iemocap} and first row of Table \ref{table:statistics}) depict a limit case, where knowledge is not transferred to different data because the same dataset is used for the RH-emo pretraining and for SER.
Therefore here we did not apply any CNN pretraining. 
Also in this special case is evident that models benefit from the use of quaternion-valued SER CNNs fed with emotional embeddings, with an average improvement of 9.74 pp among all CNN designs we tested.

\section{Ablation studies}
\label{sec:ablation}

In order to further explore the properties of our approach and to support its foundations, we performed additional experiments and ablation studies.
For these studies we applied the same experimental setup presented in Section \ref{sec:evaluation}, altering only specific details, as described below.

\subsection{Removing RH-emo components}
\label{sec:ablation_components}

\begin{figure}[!tb]
\begin{minipage}[b]{1.0\linewidth}
  \centerline{\includegraphics[width=9.5cm]{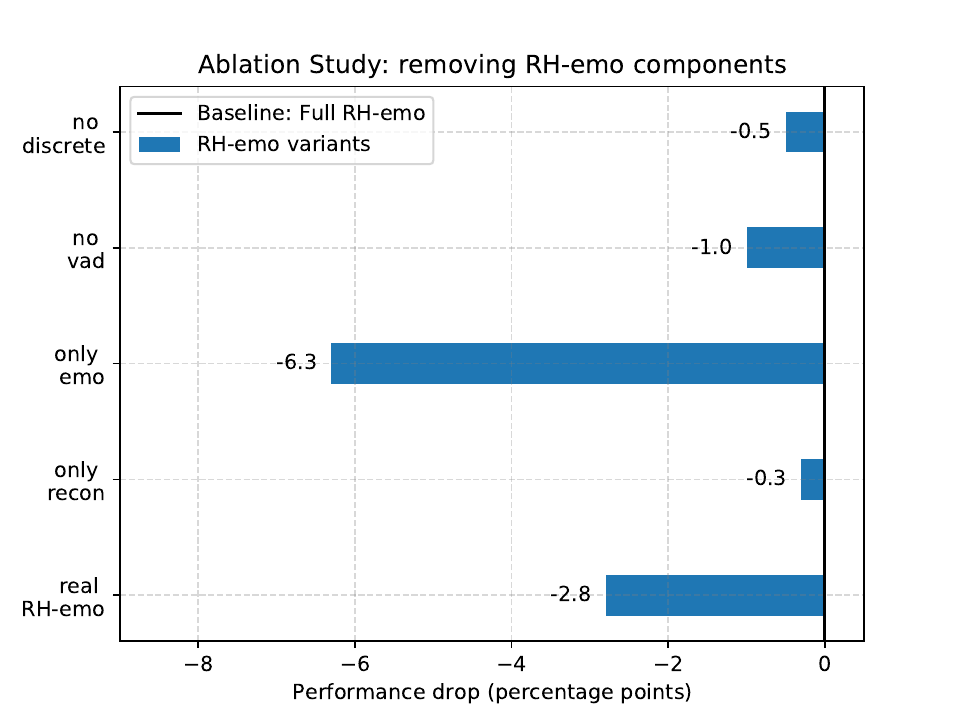}}
\end{minipage}
\caption{Ablation study results. The \textit{x} axis shows the average drop in test accuracy (among the quaternion-valued VGG16, AlexNet and ResNet-50 for all corpora) obtained with different variants of RH-emo. Each row refers to a variant of RH-emo where we removed a specific component, namely: a completely real-valued network, only reconstruction, only emotion recognition, no valence-arousal-dominance (vad) estimation, and no discrete emotion classification.}
\label{fig:ablation studies}
\end{figure}

In this study, we alter the RH-emo structure and test the emotion recognition accuracy using the embeddings generated from the modified RH-emo networks. 
We compared the full RH-emo, as described in Section \ref{sec:method}, to the following altered versions:
\begin{itemize}
    \item Real: identical to the regular network, but the decoder part is real-valued and no split activation is applied to the reconstructed output in the loss function.
    \item Reconstruction only: we removed the supervised classification branch, resulting in a completely unsupervised real-quaternion hybrid autoencoder. 
    \item Emotion only: we removed the unsupervised reconstruction branch from the network, obtaining a completely supervised and real-valued emotion classification CNN. 
    In this configuration, there are still 4 target outputs, each with a dedicated classifier (discrete emotion, valence, arousal, dominance). 
    \item Discrete emotion only: we removed the valence, arousal, and dominance classifiers, keeping only the discrete emotion classification branch.
    The rest of the network is unaltered.
    \item valence-arousal-dominance only: we removed the discrete emotion recognition branch, keeping only the branches for valence, arousal, and dominance. 
    The rest of the network is unaltered.
\end{itemize}

Figure \ref{fig:ablation studies} exposes the results of this ablation study.
In the figure, we show the mean test accuracy improvement obtained for all corpora with the quaternion-valued VGG16, AlexNet, and ResNet-50 over the real-valued baselines. 
Each row shows the results obtained feeding the quaternion-valued networks with the embeddings created with the above-described variants of RH-emo.  
These results consistently confirm the foundation of our approach.
The performance of all variants is inferior to the full RH-emo. 
In addition, we recall that the quaternion-valued CNNs fed with the emotional embeddings use a considerably lower amount of parameters.
The results point out that the unsupervised branch of RH-emo is fundamental to obtain useful embeddings, in fact, the emotion-only version, where the decoder part of RH-emo is removed, provides the most severe drop in performance compared to all variants and also the baseline.
As we expected, the quaternion-valued decoder of the actual RH-emo outperforms the completely real-valued version (by 2.8pp).
This supports our hypothesis that a quaternion-value decoder is able to create embeddings that present more suitable intra-channel correlations for the quaternion-valued CNNs.
Moreover, also here, the quaternion approach leads to faster (pre)training and less memory demand due to the lower amount of parameters.
The completely unsupervised variant (recognition-only) is conceptually similar to R2Hae \cite{parcollet2019tele}, but it relies on a convolutional design and it is applied to a different domain.
This ablation study shows that the addition of a classification branch to R2Hae provides an improvement in performance (by 0.3 pp in our case) and therefore the semi-supervision can be considered a valuable extension to R2Hae.
This ablation study also shows that the classification of emotion in the valence-arousal-dominance space is more influential in the creation of stronger embeddings.
In fact, the RH-emo variant without discrete classification provides superior accuracy compared to the discrete-only version (by 0.5 pp)
This is further supported by the fact that, as a result of an extensive grid search, we apply a stronger weight to the valence-arousal-dominance term of the loss function (the $\alpha$ term in eq.~\eqref{eq:loss}).

\subsection{Removing RH-emo pretraining and backpropagation}
\label{sec:ablation_back}

\begin{figure}[!tb]
\begin{minipage}[b]{1.0\linewidth}
  \centerline{\includegraphics[width=9.5cm]{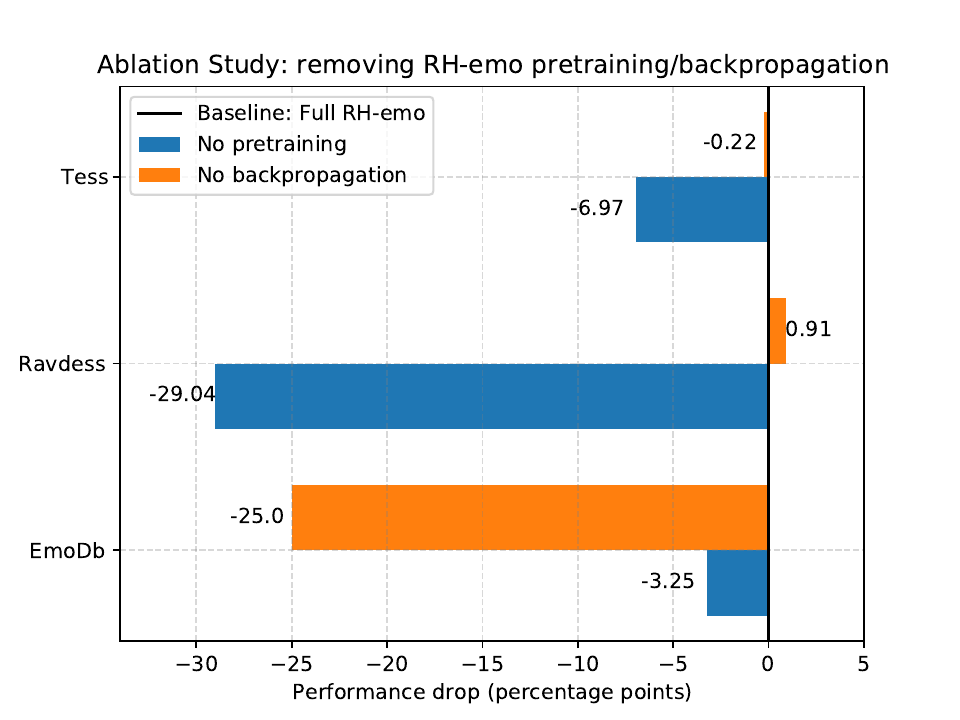}}
\end{minipage}
\caption{Ablation study results. The \textit{x} axis shows the average difference in test accuracy (among the quaternion-valued VGG16, AlexNet and ResNet-50) obtained by removing the RH-emo pretraining (blue lines) and backpropagation (orange lines).}
\label{fig:ablation_training}
\end{figure}

We performed an additional ablation study where we alter how the RH-emo weights are initialized and backpropagated during the SER training.
Figure \ref{fig:ablation_training} depicts the results of this study, showing the average difference in test accuracy per-dataset among all CNN designs.
On the one hand, we initialized the weights of RH-emo with random values while we regularly backpropagate the gradients of the RH-emo's encoder layers (blue rows). 
By doing this, we completely ignore the RH-emo pretraining and we force the QCNN network to perform end-to-end training, directly learning how to map the real-valued input spectrograms into quaternion-compatible representations to feed the QCNNs.
This approach is conceptually similar to (R2He) \cite{parcollet2019tele}.
On the other hand, we regularly initialize the weights of RH-emo with the pretrained RH-emo network, but we don't backpropagate the RH-emo layers (orange rows).
The results of this experiment strongly support the foundation of our approach.
The removal of RH-emo pretraining causes a consistent and substantial decrease in the QCNNs test performance, of 29.4, 3.25, and 6.97 pp for RAVDESS, EmoDB, and TESS, respectively.
This confirms the importance of the prior training of the RH-emo encoder, as exposed in Section \ref{sec:method}, for the development of adequate quaternion emotional embeddings.  
On the contrary, the lack of backpropagation of the RH-emo layers does not provide a consistent performance drop. 
While the performance decreases for EmoDB (25 pp ) and for TESS (0.22 pp), a narrow accuracy boost is evident for RAVDESS (0.91 pp).
Moreover, the performance difference is averagely inferior compared to the no-pretraining case.

\subsection{Reducing training data}
\label{sec:ablation_reduced}

\begin{figure}[!tb]
\begin{minipage}[b]{1.0\linewidth}
  \centerline{\includegraphics[width=9.5cm]{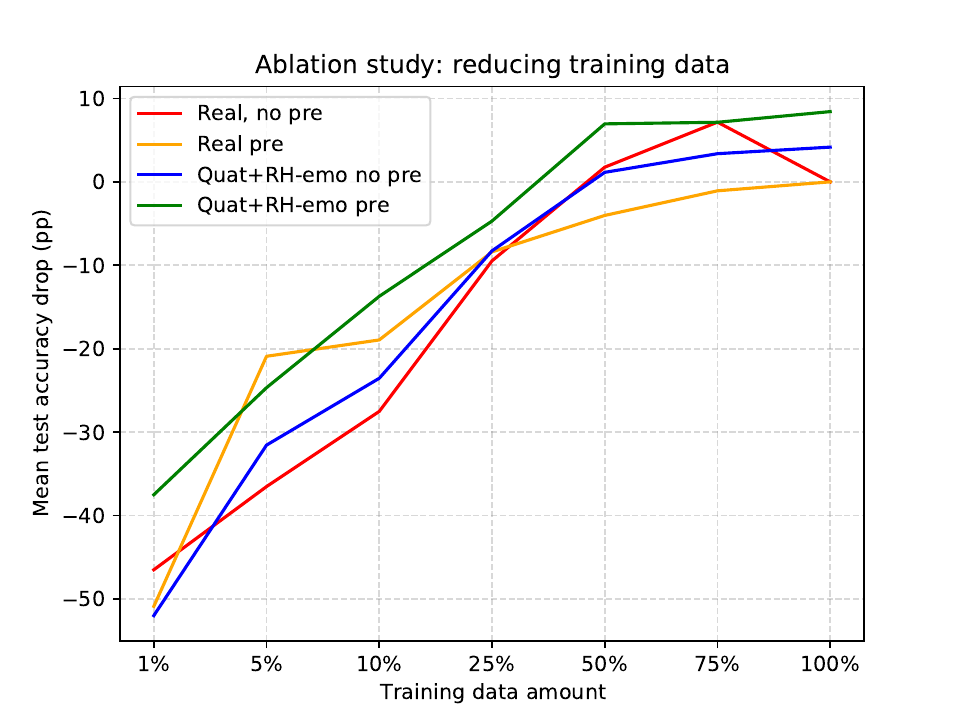}}
\end{minipage}
\caption{Ablation study results. The \textit{y} axis shows the test accuracy drop of each model, compared to the baselines that use 100\% of the training data. 
Each point in the line shows the average performance among the real-valued (red, yellow) and quaternion-valued (blue, green) VGG16, AlexNet, and ResNet-50 architectures for all corpora.
The \textit{x} axis shows the percentage of available training and validation data used.
The data reduction rates shown in the \textit{x} axis are a discrete set: we trained only on the data percentage values that are shown and not on intermediate values.
We use the full test set in all cases, in order to have a consistent performance measure also with.
}
\label{fig:ablation_reduced}
\end{figure}

As a further study, we re-trained all CNNs and QCNNs, progressively decreasing the amount of training and validation data.
The size of the test set, instead, is kept unaltered, in order to have a consistent performance measure that can be compared with the other results presented in this paper.
Figure~\ref{fig:ablation_reduced} shows the outcomes of this experiment.
Each line shows the trend of the average test accuracy among all CNN architectures, at different reduction rates of the data.
Specifically, we trained on 100\%, 75\%, 50\%, 25\%, 10\%, 5\% and 1\% of the available data.
The yellow and red lines are the baselines, respectively with and without CNN pretraining on IEMOCAP. 
Instead, the green and blue lines show the trend for the QCNNs + RH-emo, respectively with and without CNN pretraining. 

The results of this ablation study clearly point out that our method can provide consistent performance improvement even in conditions with less data.
In all cases but one (5\% of training data) our pretrained approach surpasses both real-valued baselines.
This is a convenient property for SER tasks, considering the general scarcity of emotion-labelled speech audio data.

\section{Discussion}
\label{sec:discussion}

\subsection{Resource savings}
\label{subsec:resources}

RH-emo permits to spare of a considerable amount of parameters. 
Compared to the real counterparts, the quaternion VGG16 uses the $\sim$6\% of the parameters, while the quaternion AlexNet and ResNet-50 use the $\sim$25\%.
The difference between the VGG16 and the others is due to the lack of adaptive average pooling (as described above).
Therefore, on the one hand, the use of quaternion-valued layers instead of real-valued ones permits to drop in the number of parameters by a factor of 0.25, while, on the other hand, the smaller feature dimensionality obtained with the embeddings further cuts down the number of parameters by a factor of 0.25.
This in turn permits the reduction of the model's memory requirements and training time.
In our implementation, the embedding computation happens during the training for every batch and, therefore, both the main network and the RH-emo feature extractor are loaded into the memory.
This simulates a plausible application scenario of RH-emo, where the embeddings need to be computed in real-time.
Although it is possible to pre-compute the embeddings as part of the pre-processing pipeline, further reducing the memory demand and computation time.
As regards the memory demand, in our setup the quaternion networks require on average 84.2\% of memory, compared to their real-valued equivalents.
For the VGG16 (where we don't apply average pooling) the memory demand is approximately 70\%, for AlexNet the 89\%, and for ResNet-50 the 93\%. 
Regarding the training time, the epoch duration of our quaternion networks compared to the real networks is approximately 15.9\% for VGG16, 88.1\% for AlexNet, and 162.6\% for ResNet-50. 
These outcomes show that the maximum efficiency in terms of both memory demand and computation time is obtained for VGG16, where we take advantage of the reduced dimensionality of the embeddings. 
On the other hand, the accuracy improvement for ResNet-50 comes at the cost of an increased computation time with respect to the real networks, but still reducing the model's memory demand.

\subsection{Reconstruction properties}
\label{subsec:results}

\begin{figure}[!tb]
\begin{minipage}[b]{1.0\linewidth}
  \centerline{\includegraphics[width=9.5cm]{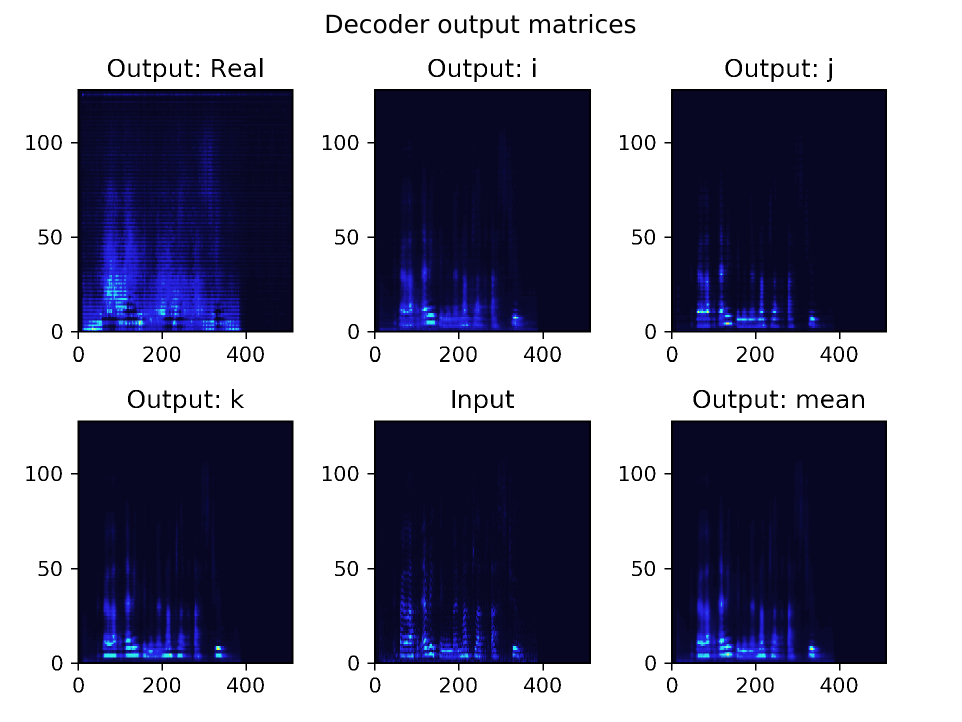}}
\end{minipage}
\caption{Example of RH-emo quaternion reconstruction of a speech spectrogram. \textit{Input} is the magnitudes-only input spectrogram, \textit{Output: real}, $\ii, \ij, \ik$ are the four output matrices of RH-emo, respectively reconstructed from the discrete emotion, valence, arousal and dominance axes of the embeddings, \textit{Output: mean} is the pixel-wise average of \textit{Output: real}, $\ii, \ij, \ik$ and is the matrix that is compared to the input in the loss function.}
\label{fig:reconstruction}
\end{figure}

Figure \ref{fig:reconstruction} shows an example of the decoder's output of the pretrained RH-emo model.
The \textit{Input} subplot is the input magnitudes-only spectrogram and the \textit{Output: mean} is the element-wise mean of the quaternion separate axes and, therefore, the actual matrix that is compared to the input in the loss function.
The sub-plots labelled as \textit{Output: real}, $\ii, \ij, \ik$ depict the separate quaternion axes, which are generated from the emotional embeddings: \textit{real} from the discrete emotion classification matrix, and $\ii, \ij, \ik$ from the valence, arousal, and dominance channels, respectively.

By comparing the \textit{Input} and the \textit{Output: mean} plots, it is evident that the reconstruction is not perfect.
While the time-wise articulation of the speech seems to be accurately reproduced, the model is not able to reconstruct in detail the most feeble harmonics of the signal.
Although it is interesting the way the different quaternion axes are differentiated.
In the real axis, the model seems to perform an operation similar to amplitude compression (obtainable, for instance, by computing the square root of the matrix), bringing up the signal's quietest portions around the speech region.
Instead, in the 3 complex axes ($\ii,\ij,\ik$) different aspects of the signal are highlighted, focusing on different harmonics and/or temporal areas.
Our intuition is that these representations may represent different ``emotional points of view" of the input speech signal.

\subsection{Limitations}
\label{subsec:limitations}
Besides the numerous advantages that our approach provides, there are also some intrinsic limitations.
The main constraint of our approach is that a pretrained RH-emo network can be used for only a fixed time scale. 
In this paper, we considered a temporal window of 4 seconds, which is well suited for most SER tasks and datasets. 
If a different time scale is needed, then a specific RH-emo has to be trained on purpose.
Another limitation is that training with an end-to-end fashion is not possible, as a pre-trained RH-emo is needed and the omission of the RH-emo pretraining stage leads to a drastic decrease in the model's performance, as shown in Section \ref{sec:ablation_back}.

\subsection{Applications and future work}
\label{subsec:applications}
The advantages provided by the combination of RH-emo and quaternion-valued networks suggest several application scenarios. 
Due to the substantial saving of trainable parameters, memory, and training time, our approach is particularly suited for situations where limited resources are available and performance can not be sacrificed.
Another useful property of RH-emo is that while the embeddings carry the necessary information to perform SER tasks (as proven by our experimental results), they also provide speaker anonymity, as it is not possible to reconstruct the input spectrogram without the RH-emo pretrained weights.
This could be exploited in situations where sensible speech data must be used for SER tasks.

The  positive  results  we obtained justify  further  investigation  of  this approach. 
An immediate research objective is to test RH-emo with different datasets, and architectures (including recurrent networks), with multiple time scales and to different tasks. 
In particular, we intend to apply the same principle of RH-emo (based on a semi-supervised autoencoder where each embedded channel is optimized for the classification of a different characteristic of an entity)
for different tasks, where a quadral representation of input data can not be directly inferred from data, as for speech emotion. 
An example of this is music genre recognition tasks, where the embedded dimensions of the autoencoder are optimized for tempo, harmonic key, spoken words, and instrument type recognition.

\section{Conclusions}
\label{sec:conclusions}
In this paper we presented RH-emo, a semi-supervised approach to obtain quaternion emotional embeddings from real speech spectrograms.
This method enables to perform speech emotion recognition tasks with quaternion-valued convolutional neural networks, using real-valued magnitudes spectrograms as input.
We use RH-emo pretrained on IEMOCAP to extract quaternion embeddings from speech spectrograms, where the individual axes are optimized for the classification of different emotional characteristics: valence, arousal, dominance, and overall discrete emotion.

We compare the performance on SER tasks of real-valued CNNs fed with regular spectrograms and 
quaternion-valued CNNs fed with RH-emo embeddings. 
We evaluate our approach on a variety of cases, using 4 popular SER datasets (IEMOCAP, RAVDESS, EmoDB, TESS) and with 3 widely-used CNN designs of increasing capacity (ResNet-50, AlexNet and VGG16).
Our approach provides a consistent improvement in the test accuracy for all datasets while using a considerably lower amount of resources. 
We obtained an average improvement of 6.01 pp for RAVDESS, 2.34 pp for EmoDB, and 0.97 pp for TESS and we spared up to 94\% of the trainable parameters, up to the 30\% of GPU memory and up to 84.1\% of training time.
Moreover, we performed additional experiments and ablations studies that confirm the properties and foundations of our approach.
The results show that the combination of RH-emo and QCNNs is a suitable strategy to circumvent the high resource demand of SER models and that our approach provides consistent performance improvement also in scenarios where the available training data is scarce. 

The positive results justify further investigation of this approach.  
An immediate research objective is to test RH-emo with different datasets, architectures (including recurrent networks), with multiple signal dimensions, and different tasks.

\section*{Acknowledgments}
The authors would like to thank the NVIDIA Applied Research Accelerator Program for the donation of an NVIDIA Quadro RTX 8000 for the project ``Quaternion Deep Learning for 3D Audio Sources''.


\end{document}